\documentclass[showpacs,amsmath,reprint,showkeys,pra,longbibliography]{revtex4-2}
\usepackage{dcolumn}    
\usepackage{bm}         
\usepackage{ifpdf}
\usepackage{amssymb,lineno,amsfonts}
\usepackage{graphicx}   
\usepackage{bbm}
\usepackage{mathrsfs}
\usepackage{upgreek}
\usepackage{epstopdf}
\usepackage{setspace}
\usepackage{hyperref}
\usepackage[matrix,frame,arrow]{xypic}
\usepackage{float}
\usepackage{soul}
\usepackage{tikz}
\usepackage{color} 
\usepackage{subfigure}
\usepackage{dsfont}
\usepackage{diagbox}
\definecolor{med-blue}{RGB}{25,25,112} 
\hypersetup{colorlinks, linkcolor={blue},citecolor={blue}, urlcolor={black}}

\newcommand{\tr}{\mathrm{Tr}}

\begin{document}
	\title{Recommender System Expedited Quantum Control Optimization}
	\author{Priya Batra}
	\email{priya.batra@students.iiserpune.ac.in}
	\author{M. Harshanth Ram}
	\email{m.harshanthram@students.iiserpune.ac.in}
	\author{T. S. Mahesh}
	\email{mahesh.ts@iiserpune.ac.in}
	\affiliation{Department of Physics and NMR Research Center,\\
		Indian Institute of Science Education and Research, Pune 411008, India}

\begin{abstract}
{Quantum control optimization algorithms are routinely used to generate optimal quantum gates or efficient quantum state transfers.
However, there are two main challenges in designing efficient optimization algorithms, namely overcoming the sensitivity to local optima and improving the computational speed.  The former challenge can be dealt with by designing hybrid algorithms, such as a combination of gradient and simulated annealing methods.  Here, we propose and demonstrate the use of a machine learning method, specifically the recommender system (RS), to deal with the latter challenge of enhancing  computational efficiency.
We first describe ways to set up a rating matrix involving gradients or gate fidelities.  
We then establish that RS can rapidly and accurately predict elements of a sparse rating matrix.  Using this approach, we expedite a gradient ascent based quantum control optimization, namely GRAPE and demonstrate the faster performance for up to 8 qubits.  
Finally, we describe and implement the enhancement of the computational speed of a hybrid algorithm, namely SAGRAPE.}
\end{abstract}

\maketitle
\noindent
\section{Introduction}
Quantum control is crucial for the trending quantum technology tasks such as quantum sensing \cite{rembold2020introduction}, scalable quantum information devices \cite{dolde2014high}, quantum simulations \cite{georgescu2014quantum}, quantum thermodynamics \cite{PhysRevA.98.012139}, quantum metrology \cite{Sekatski2017quantummetrology}, etc. In general, quantum control optimization deals with finding the best implementation of  desired quantum dynamics on a given physical hardware
\cite{brif2010control,d2021introduction}. 
Such optimization algorithms are routinely used for control tasks like unitary synthesis, state transfer, etc \cite{glaser2015training}.  
There has been a tremendous progress in this area and numerous optimization algorithms have been developed, such as gradient based algorithms \cite{khaneja2005optimal, machnes2018tunable}, variational principle based algorithms \cite{PhysRevA.68.062308,reich2012monotonically}, chopped basis optimization \cite{doria2011optimal,egger2014adaptive}, and metaheuristic algorithms \cite{fortunato2002design,zahedinejad2014evolutionary}. Lately, machine learning algorithms such as reinforcement learning (RL), have also been used for the tasks like unitary transformation \cite{an2019deep}, state preparation \cite{zhang2019does}, robust controls \cite{niu2019universal}, as well as to control non-integrable quantum systems \cite{PhysRevX.8.031086}. More recently, machine learning protocols have also been used to control quantum thermal machines \cite{khait2021optimal}.  

\begin{figure}
	\centering
	\includegraphics[trim=12cm 7.8cm 12cm 5cm,width=8cm,clip=]{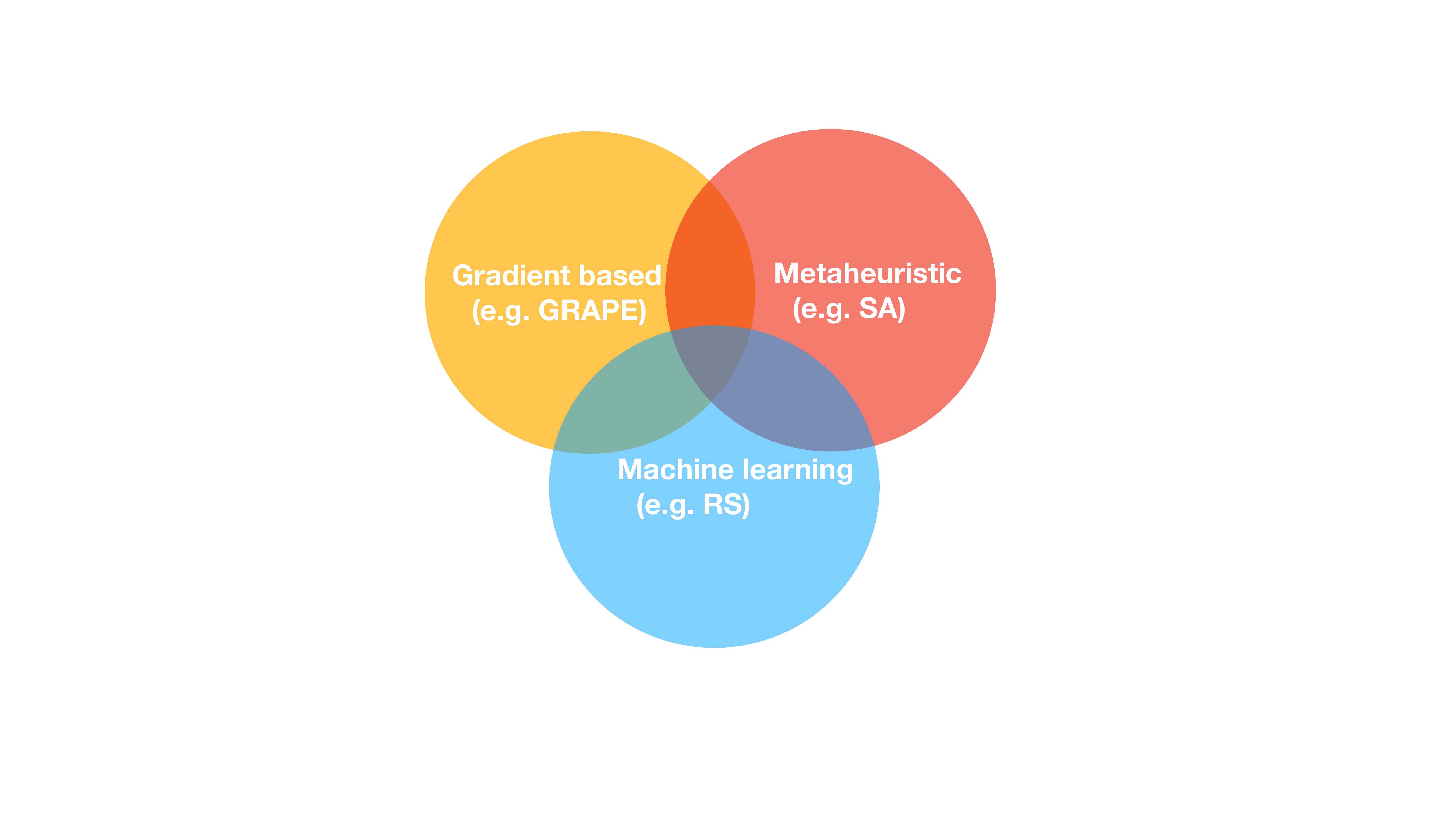}
	\caption{Expediting gradient method, metaheuristic method, as well as a hybrid of both by using machine learning.}
	\label{fig:RShybrid}
\end{figure}

An optimization method with an analytical expression for the gradient is an efficient way to find a local optimum in the parameter space.  One such class of optimization algorithms for quantum control applications is based on gradient ascent pulse engineering (GRAPE) \cite{khaneja2005optimal} and its variants (for example,
\cite{de2011second,lucarelli2018quantum,PhysRevResearch.2.013314}).   
However, it becomes problematic if there are too many sub-optimal local solutions restricting the algorithm  from reaching an optimal solution. 
On the other hand, metaheuristic algorithms can escape the local optima and therefore perform a robust search in the parameter space.  The metaheuristic  algorithms such as the Nelder-Mead simplex method \cite{fortunato2002design}, evolutionary algorithms \cite{krause1998quantum, zahedinejad2014evolutionary, bhole2016steering, PhysRevLett.114.200502}, simulated annealing (SA) \cite{ram2021robust}, etc. have been successfully  adapted for quantum control optimization. 
However, unlike the gradient methods, the metaheuristic algorithms often suffer from slow convergence.  
To overcome this issue, recently there has been a proposal to combine SA, a metaheuristic method with GRAPE, a local search method, to realize the hybrid SAGRAPE algorithm  \cite{ram2021robust}.   Despite this progress, there remains the problem of the poor computational efficiency of quantum optimization algorithms.  This problem becomes severe as the parameter space grows exponentially with the size of the quantum system.
This is where the capabilities of machine learning can be useful (see Fig. \ref{fig:RShybrid}).

In the field of machine learning, the recommender system (RS) algorithm is widely used to recommend products to consumers \cite{lu2012recommender,bobadilla2013recommender}. Here we propose RS assisted speed up of GRAPE and SAGRAPE. The most time-expensive task in GRAPE is the calculation of gradients for every segment of a control sequence,  which involves exponentiation of matrices.  We show that, given a set of exactly calculated gradients, RS can accurately and rapidly  predict the remaining ones.  Thus, we can significantly improve the time-efficiency of the optimization algorithm.
SA involves a large number of function evaluations in the neighborhood of an iterative solution, part of which can also be delegated to RS. 
Here we demonstrate a significant time advantage of RS expedited GRAPE as well as SAGRAPE, without sacrificing their convergence.
 
This article is arranged as follows. In Sec. \ref{sec:QCT}, we introduce the quantum control problem along with two optimization algorithms, namely GRAPE and SA. In Sec. \ref{sec:RS} we introduce RS. In Sec. \ref{sec:hybrid}, we present the RS expedited GRAPE and SAGRAPE, and discuss their performances. The article ends with a summary and outlook in Sec. \ref{sec:summary}.

\section{Quantum Control Optimization}
\label{sec:QCT}
We consider a quantum system with a 
constant internal Hamiltonian $H_0$ and $m$ 
control Hamiltonians $\{H_j\}$, such that 
the time-dependent Hamiltonian is given by
\begin{equation}
	\label{ham}
	H(t) = H_0+ \sum_{j=1}^{m}\omega_j(t) H_j.
\end{equation}
Here $\omega_j(t)$ is the time-dependent strength of the $j$th control Hamiltonian $H_j$. 
For a closed system undergoing Schr\"odinger  evolution for the time duration $T$, the corresponding unitary is given by
\begin{equation}
	\label{unitary}
	U(T) = D e^{-i \int_{0}^{T} H(t) dt}.
\end{equation}
Here $D$ is the Dyson time ordering operator and we have set $\hbar = 1$.
Considering the difficulty in evaluating the propagator of a general time-dependent Hamiltonian, we discretize the control function $\omega_j(t)$ by dividing it into $N$ segments, each with a constant amplitude $\omega_{j,k}$ and duration $\delta = T/N$.
The corresponding unitary for the $k$th segment would be 
\begin{equation}
U_{k} = \exp\left[-i \delta \left(H_0 + \sum_{j=1}^{m} \omega_{j,k} H_j\right) \right].
\end{equation}
The total time propagator $U(T)$ for the entire control sequence can be expressed as the product of segment unitaries
\begin{equation}
U(T) = U_N U_{N-1} \cdots U_2 U_1.
\end{equation}

In this article, we mainly focus on unitary synthesis, although same methods can be adapted for state preparation.
Thus, the optimization procedure is aimed at achieving a specific target unitary $U_t$ by numerically generating the control sequence $\{\omega_{j,k}\}$.  Optimization function is given by the gate fidelity 
\begin{align}
F =  \left\vert \tr \left[U^\dagger(T)U_t \right]
\right\vert^2,
\end{align}
which is the overlap of target unitary $U_t$ with the evolved unitary $U(T)$.
An improved optimization function would maximize the fidelity while minimizing  resources.  Typically, it amounts to minimizing the power consumption of control fields.  Therefore the modified optimization function $J$ can be cast as follows 
\begin{align}
J = F -  \sum_{j,k} \lambda_j \omega_{j,k}^{2},
\end{align}
where $\lambda_j$ are the scalar penalty parameters.
In the following, we discuss two optimization methods, first a gradient method and the second a metaheuristic method.

\subsection{Gradient Ascent Pulse Engineering (GRAPE) \label{grape}}
The GRAPE algorithm has following steps:
\begin{itemize}
	\item Start with a random control sequence $\{\omega_{j,k}^{(0)}\}$.
	\item Forward propagate the initial unitary operator $U_0 = \mathbbm{1}$ till $k$th segment, i.e.,
	$X_k = U_k U_{k-1} \cdots U_1 U_0$.
	\item Backward propagate the target unitary $U_t$ till $k$th segment, i.e., $P_k = U_{k+1}^{\dagger}U_{k+2}^{\dagger} \cdots U_{N}^{\dagger}U_t$. 
	\item Calculate the gradient for each segment using the first-order (in the norm $||\delta H(t)||$) expression \cite{khaneja2005optimal}
	\begin{align}
	g_{j,k} =  2 \mathrm{Re} \left(-i\delta \tr \left[
	P_k^\dagger H_j X_k	\right] \tr \left[X_k^\dagger P_k \right] \right) .
	\label{eq:gradgrape}
	\end{align}
    \item
	Update control amplitudes in the direction of gradients, i.e., $\omega_{j,k} \rightarrow \omega_{j,k} + \epsilon g_{j,k}$, where 
	$\epsilon$ is the step size. 
\end{itemize}  
GRAPE is remarkable to have such a simple analytical form for the gradient function. 

In a practical scenario, no physical hardware is perfect.  For example, a control field is typically associated with a distribution of amplitudes around the nominal value \cite{khaneja2005optimal}.  
We need to have a robust quantum control even with such imperfect hardware.  
To this end, one minimizes the overall cost function 
$1-\overline{J} =1- \sum_{i} p_i J_i$
obtained by summing over the costs $1-J_i$ of individual elements in the distribution with respective probabilities $p_i$.  



\subsection{Simulated Annealing (SA) \label{sec:sa}}
In metallurgy, annealing involves heating a material to high temperatures followed by slow cooling, to allow the material  reach a stable crystalline form by finding its ground state.  The same idea is adopted in the numerical procedure namely, simulated annealing (SA) \cite{aarts1987simulated,bertsimas1993simulated}.  Given an optimization problem, SA starts with a high-temperature configuration, wherein even nonoptimal solutions are selected with a certain probability.  As the iterations pass, the temperature parameter is gradually reduced, and  optimal solutions are increasingly favored.  This stochastic process allows the algorithm to overcome the local minima and reach the global minimum.

The steps for SA are as follows \cite{ram2021robust}:
\begin{itemize}
    \item Start with a random control sequence $\{\omega_{j,k}^{(0)}\}$.  Set the temperature to a high value $T^0$.
    \item In the $i$th iteration, near the current solution $\{\omega_{j,k}^{(i)}\}$, determine the control sequence $\{\omega_{j,k}'\}$ with the best fidelity $\overline{F}(\{\omega_{j,k}'\})$ among a random set of neighbourhood points.
    \item  If
    $\Delta \overline{F}^i = \overline{F}(\{\omega_{j,k}'\})-\overline{F}(\{\omega_{j,k}^{(i)}\}) \ge Q^i$, where \\
    $Q^i = -\mathrm{min}
    \left[
    1, T^i \exp\left(
    \Delta \overline{F}^i/T^i
    \right)
    \right]$ then \\ $\{\omega_{j,k}^{(i+1)}\}= \{\omega_{j,k}'\}$;  else,  $\{\omega_{j,k}^{(i+1)}\}= \{\omega_{j,k}^{(i)}\}$.
    \item After $i$th iteration, set the  temperature to a reduced value $T^{(i+1)} = \gamma T^{(i)}$, where $\gamma < 1$ controls the cooling rate. 
\end{itemize}
The above steps are iterated until the desired optimization function is reached or the maximum number of iterations are completed.
Note that, for higher temperatures the algorithm may take a new solution $\{\omega_{j,k}'\}$ even if its fidelity is lower than the current solution $\{\omega_{j,k}^{(i)}\}$.  This stage is known as \textit{exploration}.  As temperature goes down, the algorithm gradually switches to the \textit{exploitation} mode and it looks for solutions that are either better or slightly inferior than the current solution.   The combination of exploration and exploitation helps SA to escape local optima and travel towards the global optimum.  

In our previous work, we had combined SA with GRAPE to form the hybrid SAGRAPE algorithm and demonstrated its superior convergence \cite{ram2021robust}. In the following, we first describe recommender system, a particular type of machine learning technique, and then explain how it can be used to expedite GRAPE as well as SAGRAPE.

\section{Recommender System (RS)}
\label{sec:RS}
 Collaborative filtering is one of the most popular types of RS that is based on experience of any particular consumer along with relative preferences among all consumers \cite{herlocker2004evaluating,su2009survey}. 
Here we use the matrix factorization algorithm for collaborative filtering  \cite{singh2008unified,koren2009matrix}. It involves setting up a database ${\cal R}$ in the form of a rating matrix, wherein each row represents a particular consumer and each column represents a particular product that is being recommended \cite{bobadilla2013recommender,Batra2021efficient}.  The rating matrix can be decomposed in terms of latent vectors of the same dimension $f$, known as the number of features.  Let the parameter vector $\Theta^{(i)} \in \mathbb{R}^f $ and the feature vector $\Phi^{(j)}\in \mathbb{R}^f $ encode latent vectors in real space $\mathbb{R}^f$ for $i$th consumer and $j$th product. The predicted rating is then modeled by  scalar products
\begin{equation}
r_{i,j} = \Theta^{(i)} \cdot \Phi^{(j)}. 
\end{equation}      

\begin{figure}
	\centering
	\includegraphics[trim=0cm 0cm 0cm 0cm,width=8cm,clip=]{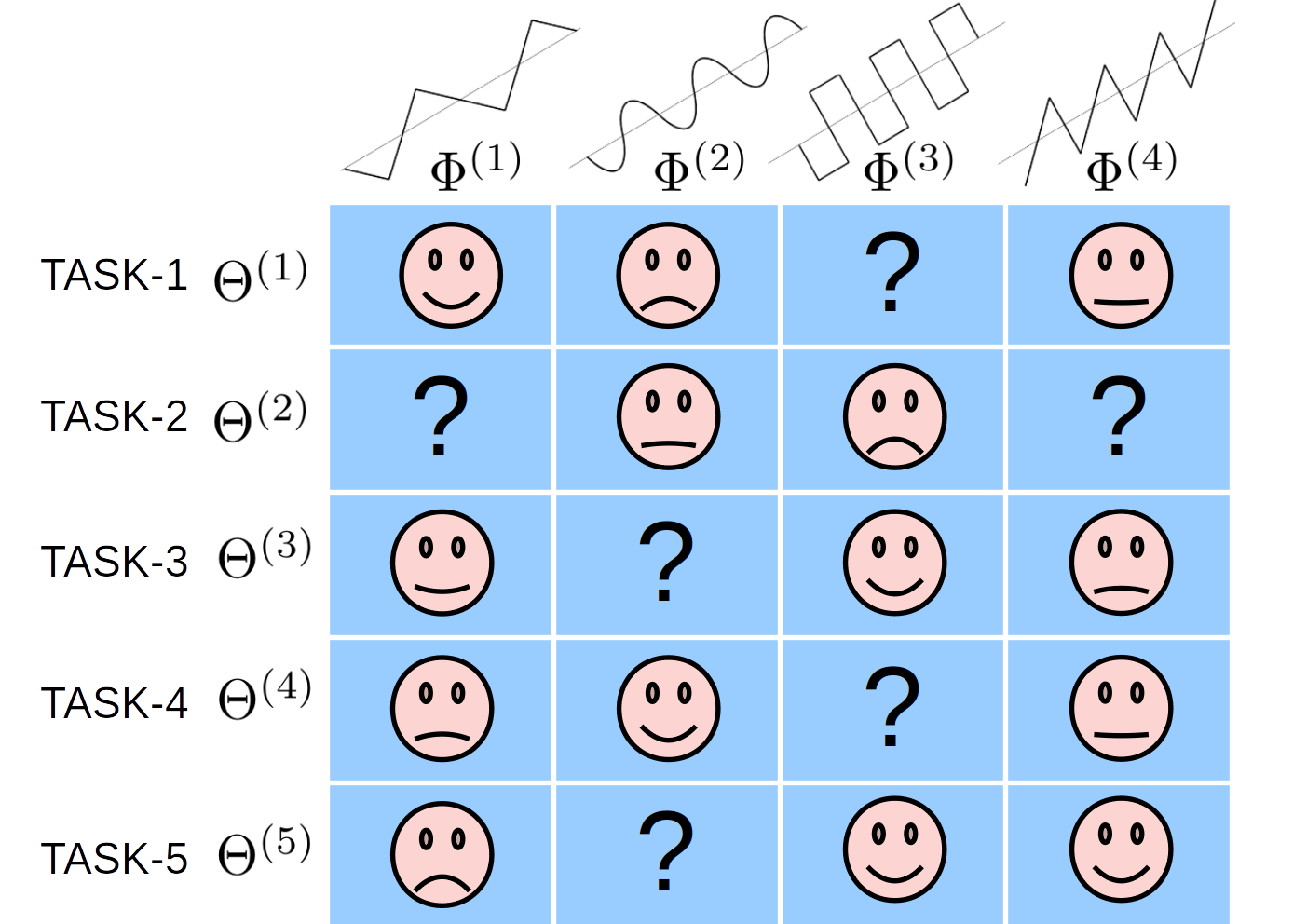}
	\caption{Illustrating an example application of RS.  Here the rating matrix corresponds to different levels of feasibility of various tasks with certain available control fields. The goal of RS is to predict the unknown ratings.}
	\label{fig:rsqc}
\end{figure}

Depending on the problem at hand, one can specify products, users, as well as ratings, and accordingly set up the rating matrix.  One such example, for a hypothetical quantum control problem of executing certain specific tasks with various available control fields, is illustrated in Fig. \ref{fig:rsqc}.  In this example, tasks are users, control fields are products, and the ratings are different levels of  feasibility of implementing tasks.  The job of RS is to efficiently predict the unknown ratings.


Given a sparse rating matrix ${\cal R}$, where missing elements correspond to unknown preferences, our goal is to fill those missing elements with the help of the collective information embedded in the overall database. Let $\kappa = {(i,j)}$ be the elements in the rating matrix for which actual ratings ${\cal R}_{i,j}$ are known. The discrepancy between predicted rating $r_{i,j}$ and actual rating ${\cal R}_{i,j}$ is quantified by the function $K_0$ as 
\begin{equation}
K_0 = \sum_{(i,j)\in \kappa } (r_{i,j} - {\cal R}_{i,j})^2.
\end{equation}

Here, the objective is to minimize the cost function $K_0$. Generally two regularization parameters $(\Lambda_{\Theta},\Lambda_{\Phi})$ are used to avoid over-fitting, so that the altered cost function can be written as
\begin{equation}
	K = \frac{K_0}{2} 
	+ \frac{\Lambda_{\Theta}}{2} \sum_{i=1}^{m} \Vert \Theta^{(i)} \Vert + \frac{\Lambda_{\Phi}}{2} \sum_{j=1}^{n} \Vert \Phi^{(j)} \Vert.
\end{equation}
The latent vectors $\Theta^{(i)}$ and $\Phi^{(j)}$ are then determined by minimizing the cost function using any standard minimization algorithm.  In our case, we use the Polack-Ribiere flavour of conjugate gradients to compute search directions, and a line-search using quadratic and cubic polynomial approximations for this purpose \cite{refId0,press1992numerical,coficode}.



\section{RS enhanced GRAPE and SAGRAPE}
\label{sec:hybrid}
We shall now explain how we can use RS to speedup GRAPE as well as SAGRAPE.  For the sake of clarity, we use the context of quantum control of spin-dynamics via nuclear magnetic resonance (NMR).  To study RS enabled speedup, we consider a spin system of $M = 2+n$ on-resonant heteronuclear qubits, and construct a controlled-NOT (CNOT) gate on the first two qubits which are coupled by an indirect spin-spin interaction of strength ($J_{12}$).  For simplicity, we treat the remaining $n$ qubits as noninteracting spectator spins, which contribute to the dimension of the overall Hilbert space, but not to the complexity of the CNOT gate.  This helps us to demonstrate the efficiency of RS expedited algorithms for a varying size of qubit register with same gate complexity.
The internal NMR Hamiltonian for the system in the multiply-rotating frame is given by
\begin{equation}
H_0 = 2\pi J_{12} I_{z}^1 I_{z}^2, 
\end{equation}
where $I_{z}^i$ represents the $z$-component of the $i$th spin operator.  For the $k$th segment of the $n+2$ channel control sequence, the total Hamiltonian 
\begin{equation}
H_k = H_0 + \sum_{j=1}^{2+n}
\omega_{x,j,k} I_x^j
+\omega_{y,j,k} I_y^j.
\end{equation}
Here $\omega_{x,j,k}$ and $\omega_{y,j,k}$ are the $x$ and $y$ components of radio frequency (RF) field on the $j$th nuclear species in $k$th segment of the control sequence.  In practice however, there exists a spatial RF inhomogeneity (RFI) of amplitudes over the volume of the sample. One usually models RFI with an $L$-point distribution and associates a scaling factor $\{\xi_{j,l}\}$ with respective probabilities $\{p_{j,l}\}$. Thus, the Hamiltonian for the $k$th control segment with the $l$th RFI element is
\begin{equation}
H_k^{(l)} = H_0 + \sum_{j=1}^{2+n}
\xi_{j,l}\omega_{x,j,k} I_x^j
+\xi_{j,l}\omega_{y,j,k} I_y^j.
\label{eq:hamiltonian}
\end{equation}
As explained in the last part of Sec. \ref{grape}, the overall cost function $1-\overline{J} = 1-\sum_{l} p_l J_l$ is obtained by the weighted sum of cost functions of all the individual elements in the RFI distribution.  In the following, we describe how we can incorporate RS to expedite GRAPE.

\subsection{RS expedited GRAPE (RSGRAPE)}
Let us first consider the GRAPE algorithm for the NMR context described above.  
As explained in Sec. \ref{grape}, the $x(y)$ gradients for the $j$th channel in $k$th segment with $l$th RFI element are expressed by the explicit form of Eq. \ref{eq:gradgrape}, i.e.,
\begin{align}
g_{x(y),j,k,l} &=   2 \mathrm{Re} \left(-i\delta \tr \left[
P_{k,l}^\dagger I_{x(y)}^j X_{k,l}	\right] \tr \left[X_{k,l}^\dagger P_{k,l} \right] \right).
\label{eq:gradgrapenmr}
\end{align}
Here $X_{k,l}$ and $P_{k,l}$ are respectively the forward and backward propagators for the $k$ segment with $l$th RFI element.
In the traditional GRAPE algorithm, one evaluates all the gradients and then calculates the update values for the control amplitudes as described in Sec. \ref{grape}.
This is the heart of the algorithm and involves evaluating a large number of propagators via matrix exponentiation.  Accordingly, this is the bottleneck for numerical efficiency, especially for the larger number of qubits.
Here comes the application of machine learning.  Instead of evaluating all the gradients, we only need to evaluate a fraction of the gradients, form a rating matrix, and then let RS predict the rest of the gradients. In the language of RS, we treat each segment as a consumer and each RF amplitude (corresponding to indicies ${x(y),j,l}$) as a product. The corresponding rating matrix is in the form of the TABLE \ref{tab:gradrating}.

\begin{table}
\begin{tabular}{|p{3cm}|c|c|c|c|}
\hline 
RFI & \multicolumn{4}{c|}{$l = 1$ $\cdots$ }  
\\
\hline
Qubits & \multicolumn{2}{c}{j=1 $\cdots$ } & \multicolumn{2}{|c|}{j=2+n} 
 \\
\hline
\diagbox[width=\dimexpr \textwidth/8+12\tabcolsep\relax, height=1cm]{segments (k)}{phases} & x & y & x & y 
       \\
\hline
~~~~~~~1 & $g_{x,1,1,1}$ & $g_{y,1,1,1}$ & ? & $g_{y,2+n,1,1}$ 
\\
\hline
~~~~~~~2 & $g_{x,1,2,1}$ & ? & $g_{x,2+n,2,1}$ & ? 
\\
\hline
~~~~~~~3 & $g_{x,1,3,1}$ & $g_{y,1,3,1}$ & ? & $g_{y,2+n,3,1}$ 
\\
\hline
~~~~~~~4 & ? & $g_{y,1,4,1}$ & $g_{x,2+n,4,1}$ & ? 
\\
\hline
~~~~~~~~$\vdots$ & $\vdots$ & $\vdots$ & $\vdots$ & $\vdots$ \\
\hline
\end{tabular} 
\caption{The rating matrix for RSGRAPE.  Here rows correspond to various segments ($k=1,2,\cdots, N$) and columns correspond to the $x(y)$ RF amplitudes  on all the heteronuclear qubits ($j=1,2,\cdots,2+n$) with various RFI distribution elements ($l=1,2,\cdots, L$).
The elements $g_{x(y),j,k,l}$ are the gradients, and the goal of RS is to predict the unknown elements (indicated by `?').
\label{tab:gradrating}
}
\end{table}   
   
We first study the dependence of 
RSGRAPE on the sparsity of the rating matrix.  To this end, we generate a two-qubit CNOT gate while varying the sparsity of the rating matrix from 0\% to 90\%.  In each case, we monitor, at the end of 500 iterations, the final infidelity and the time advantage 
\begin{align}
\Gamma(\mbox{RSGRAPE}) &=  \tau(\mbox{RSGRAPE})/\tau(\mbox{GRAPE}),
\end{align}
the ratio of computational times of RSGRAPE and  GRAPE.  

\begin{figure}
	\hspace*{-0.3cm}
  	\includegraphics[trim=1cm 5.9cm 1.5cm 5.5cm,width=9cm,clip=]{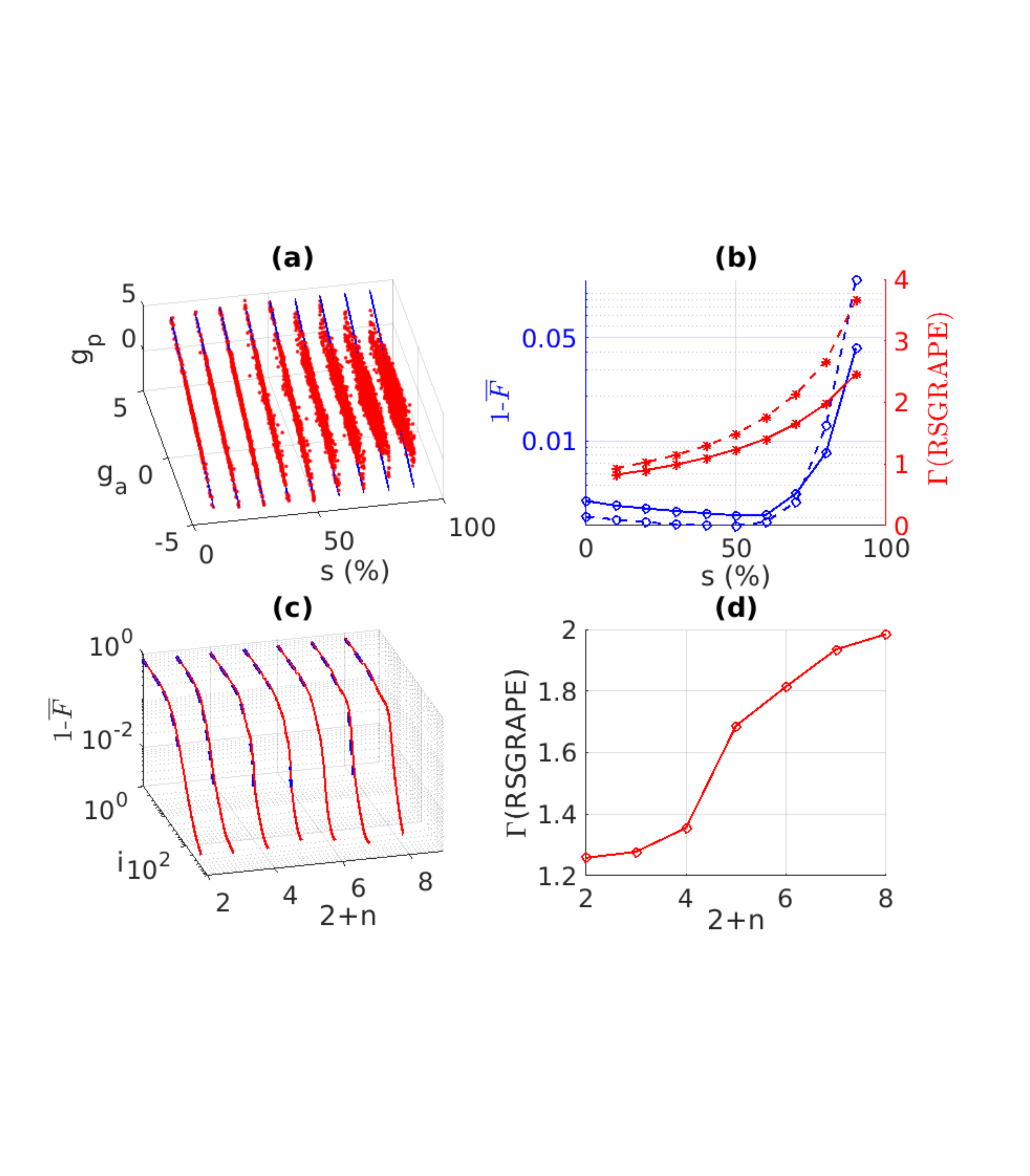}
	\caption{(a) Exact $g_a$ and predicted $g_p$ gradients vs sparsity (s in \%). 
	(b) Infidelity  $1-\overline{F}$ (blue) and time advantange $\Gamma(\mbox{RSGRAPE})$ (red) vs sparsity ($s$ in \%) of a CNOT gate generated using RSGRAPE in two (solid line) and four (dashed line) qubit systems.  
	(c) Infidelity $1-\overline{F}$ vs number $i$ of iterations varying from 1 to 500 and for size $2+n$ of the qubit register varying from 2 to 8.  Here solid-red and dashed-blue lines correspond to RSGRAPE and GRAPE  algorithms.  
	(d) $\Gamma(\mbox{RSGRAPE})$ vs size $2+n$ of the qubit register.  Here (c) and (d) are obtained with a sparsity value of 50 \%.}
	\label{fig:grapers}
\end{figure}

We now demonstrate RSGRAPE for generating an optimal control sequence implementing a CNOT gate on the first two qubits of the $2+n$ qubit system.  In our analysis we  have varied $2+n$ from $2$ to $8$ and in each case generated an optimal control sequence with $N=200$ segments.  For RS prediction we used latent vectors of dimension 10.  RFI is modeled by $L=5$ point distribution with $\xi_{j,l}\in [0.8,0.9,1.0,1.1,1.2]$ and 
uniform probability $p_{j,l} = 0.2$.
Firstly, we vary the sparsity (s) from $10\%$ to $90\%$ and compare the gradients for the standard GRAPE ($g_a$) vs gradients predicted by RS ($g_p$) as shown by red dots in Fig. \ref{fig:grapers} (a). 
For reference, also shown are the ideal expected curves $g_a=g_p$ (in blue lines in Fig. \ref{fig:grapers} (a)).
It's clear that RS is able to predict the gradients quite efficiently, especially for sparsity values below $60 \%$.  For larger sparsity values, the RS predictions become increasingly inaccurate and therefore worsens the convergence of the algorithm.

Now, we iterate (for $i = 1,\cdots,500$) GRAPE (as explained in Sec. \ref{grape}) as well as RSGRAPE (as explained above) algorithms.  In Fig. \ref{fig:grapers} (b), we plot the final infidelity $1-\overline{F}$ (blue curves) as well as time advantage $\Gamma(\mbox{RSGRAPE})$ (red curves) vs sparsity ($s$) for two (solid curves) and four (dashed curves) qubit systems.  Here for reliable analysis, all the data points are obtained by averaging $10$ independent trials each starting from a random initial guess. We find that the final infidelity remains low ($<0.005$) till about $60\%$ sparsity, and increases afterwards.  Also, we achieve time advantages by over a factor of 1.5 for both two and four qubits systems. 

We now set the sparsity of the rating matrix to 50\%, meaning only half the number of gradients randomly selected out of the total $2N (2+n)L$ elements need to be evaluated using Eq. \ref{eq:gradgrapenmr}. The rest of the gradients in each iteration are predicted by the RS algorithm.
Again, for reliable analysis, we average the results of 15 independent trials of RSGRAPE algorithm each starting from a random initial guess.  For comparison, we also carryout the standard GRAPE algorithm in each case and monitor the time advantage
$\Gamma(\mbox{RSGRAPE})$.
Fig. \ref{fig:grapers} (c) shows the infidelity ($1-\overline{F}$) versus iteration number $i$ for various sizes $2+n$ of the qubit register.  It is clear that the convergence of RSGRAPE is not compromised compared to the standard GRAPE algorithm, despite only 50\% of the gradients being exactly evaluated.  In all the cases, the infidelity was well below 0.01.  

Fig. \ref{fig:grapers} (d) shows the time-advantage for various sizes of the qubit register.  For small qubit registers (up to 4 qubits), the advantage is above 20 \%, while for larger registers (for 8 qubits), we find almost 100\% time advantage.  This is because the RS overhead is dependent on (i) the dimension of  the rating matrix, which increases only linearly with the size of qubit register and (ii) the dimensions of latent vectors.
Therefore, as the complexity of  GRAPE algorithm increases exponentially with the size of the qubit register, the RS overhead becomes increasingly insignificant, and the time-advantage improves.  However, for 50 \% sparsity, $\Gamma(\mbox{RSGRAPE})$ remains bounded by a factor of 2.

\subsection{RS expedited SAGRAPE (RSSAGRAPE)}
We now explain the RS expedited hybrid SAGRAPE algorithm. As explained in Sec. \ref{sec:sa}, an important step in SA in every iteration is to scan the neighborhood points of the current solution.  We add random deviation functions to the current control sequence (of each heteronuclear qubit) to obtain a set of neighborhood points.
Since one needs to scan a large number of neighbourhood points, this step forms a bottleneck in the standard SAGRAPE algorithm.  This is where RS can bring about a significant speedup.  

The entire control sequence of current iteration can be represented in the matrix form 
$\Omega^{(i)} \equiv \{\omega^{(i)}_{x(y),j,k}\}$ of dimension $N\times 2(2+n)$.
In our method, we first select a set of $Q$ spline vectors $\{s_{q}\}$, each of dimension $N\times 1$. We also choose a set of $M$ random scaling vectors $\{c_m\}$, each of size $1\times 2(2+n)$.
We now setup the neighborhood points by adding the deviation function $w^{(i)} s_{q} \cdot c_m$, i.e.,
\begin{equation}
\Omega_{q,m} =  
\Omega^{(i)} + w^{(i)} c_m \cdot s_{q},
\end{equation}
where $w^{(i)}$ is a scalar weight parameter which can be gradually reduced with iteration number to shrink the neighbourhood region.  Now, we determine the average fidelity
$\overline{F}_{q,m}$ and form the rating matrix (see TABLE. \ref{tab:sagrape}).  In the RS expedited algorithm, we don't need to evaluate all the elements of the rating matrix, but only a set of randomly selected elements.  Rest of the elements are efficiently predicted by RS. The sequence $\Omega_{q,m}$ corresponding to the maximum fidelity $\overline{F}_{q,m}$ is then passed to the SA algorithm for comparing with the threshold function as explained in Sec. \ref{sec:sa}.


\begin{table}
\begin{tabular}{|c|c|c|c|c|}
\hline
\diagbox[width=\dimexpr \textwidth/7+4\tabcolsep\relax, height=1cm]{spline ($s_q$)}{scaling ($c_m$)} & $c_1$ & $c_2$ & $c_3$ & $c_4\cdots$ 
       \\
\hline
$s_1$ & $\overline{F}_{1,1}$ & $\overline{F}_{1,2}$ & ? & $\overline{F}_{1,4}$  \\
\hline
$s_2$ & $\overline{F}_{2,1}$ & ? & $\overline{F}_{2,3}$ & ? \\
\hline
$s_3$ & $\overline{F}_{3,1}$ & $\overline{F}_{3,2}$ & ? & $\overline{F}_{3,4}$ \\
\hline
$s_4$ & ? & $\overline{F}_{4,2}$ & $\overline{F}_{4,3}$ & ? \\
\hline
$\vdots$ & $\vdots$ & $\vdots$ & $\vdots$ & $\vdots$ \\
\hline
\end{tabular} 
\caption{The rating matrix for the RSSAGRAPE algorithm.  Here rows correspond to various spline functions $s_{q}$ and columns correspond to the scaling factors $c_m$.  Each element of rating matrix is the fidelity $\overline{F}_{q,m}$ of the neighbourhood point $(q,m)$ obtained from the current solution by adding the deviation function  $w^{(i)}  s_{q} \cdot c_m$. 
\label{tab:sagrape}
}
\end{table}

\begin{figure}[b]
	\centering
	\includegraphics[trim=0cm 0cm 1cm 0cm,width=9cm,clip=]{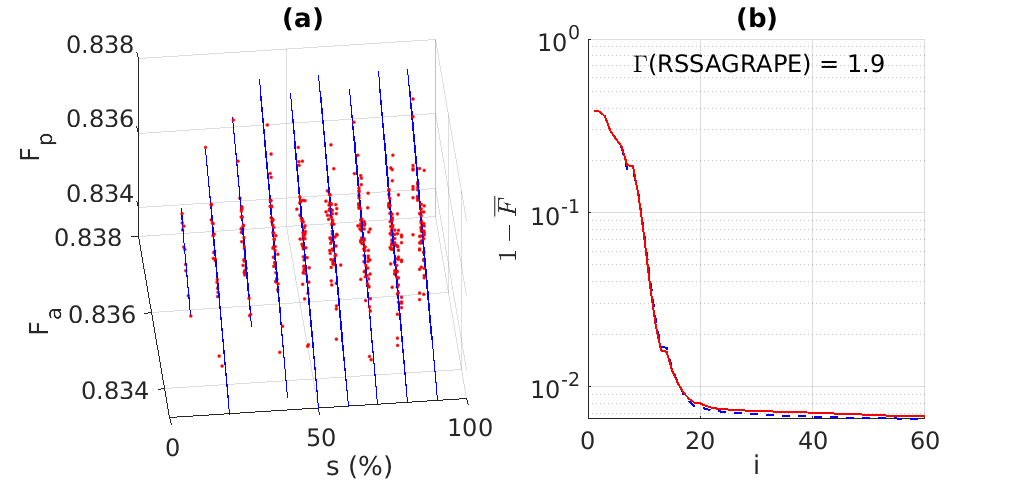}
	\caption{(a) Exact ($F_a$) and predicted ($F_p$) fidelities of random neighbourhood points versus sparsity ($s$ in \%) in a particular SA iteration.
	(b) Infidelity $1-\overline{F}$ vs number of iterations $i$ for a 2 qubit system with GRAPE (dashed blue line) or RSSAGRAPE (solid red line).  In (b) RS was carried out with sparsity $s=50$ \%.}
	\label{hybrid}
\end{figure}

Based on the procedure explained above, we now employ RSSAGRAPE to generate a CNOT gate on a two-qubit system, using five iterations of GRAPE after every iteration of SA.  
We used $N=200$ segment control sequence and scanned 100 neighbourhood points (using $Q=10$ spline functions and $M=10$ scaling vectors) in each SA iteration. 
We again set the dimension of the latent vectors to 10.  The red dots in Fig.  \ref{hybrid}(a) correspond to the exact fidelities ($F_a$) plotted against the predicted  fidelities ($F_p$) for a set of random neighborhood points at various values of sparsity $s$.  
The blue lines corresponding to expected distribution $F_p = F_a$ are also shown for reference.  Evidently, we see a good correlation of the predicted fidelities with the exact values, thus confirming the effectiveness of RS, especially for lower sparsity values.

We now set the sparsity of the rating matrix to 50\%, meaning only half the neighborhood points are evaluated exactly, while the remaining ones are predicted by RS.   
Fig. \ref{hybrid}(b) displays the  infidelity $1-\overline{F}$ (averaged over 10 independent trials each starting with a random initial guess) plotted versus iteration number $i$ for SAGRAPE (dashed blue line) as well as RSSAGRAPE (solid red line).
It is clear from the average infidelity trajectory that the convergence is not sacrificed by the partial prediction of the rating matrix by RS.  Furthermore, when compared with the SAGRAPE algorithm, we found a time advantage 
\begin{align}
\Gamma(\mbox{RSSAGRAPE})=
\frac{\tau(\mbox{SAGRAPE})}{\tau(\mbox{RSSAGRAPE})} = 1.9,
\end{align}
meaning the RS enhancement has almost doubled the speed of the SAGRAPE algorithm.

\section{Summary and outlook}
\label{sec:summary}
Machine learning techniques are increasingly being utilized in almost every field of science. Here we use recommender system (RS), a type of machine learning technique to enhance the efficiency of quantum control algorithm, particularly  a gradient method (GRAPE) and  a meta-heuristic method (simulated annealing (SA)). Because of the analytical form for gradients, GRAPE is a powerful tool, but it suffers from two problems.  Firstly, being a local search method, GRAPE is sensitive to local minima.  Secondly, evaluating gradients for all the constant-Hamiltonian segments is a computationally intensive task.  
The former can be overcome by a hybrid algorithm such as SAGRAPE, which had previously been demonstrated \cite{ram2021robust}.  Here we address the latter issue by employing  RS to efficiently predict gradients and thereby to remarkably speed up the GRAPE algorithm.  We demonstrated the RSGRAPE algorithm for up to eight qubits.
Going further, we incorporated RS even in SA, for rapid evaluation of a large set of random neighborhood points.  Finally, by expediting both SA and GRAPE simultaneously, we demonstrated almost doubling the speed of SAGRAPE.


The entirely different approaches of using RS in GRAPE and SA exemplifies the flexibility and freedom of incorporating RS in quantum control problems. Note that the particular approaches we have used are not unique. One can think of  different ways of encoding consumers and products to set up a rating matrix for implementing RS. The generality of RS approach should allow its application in conjunction with other gradient methods such as BFGS \cite{de2011second}, function-space control \cite{lucarelli2018quantum}, etc.
RS can also be used to enhance other meta-heuristic algorithms as well as global search methods such as genetic algorithm \cite{bhole2016steering}.  We believe, the present work encourages further applications of machine learning protocols in quantum information tasks.



\section*{Acknowledgements}
PB acknowledges support from the Prime Minister’s Research Fellowship (PMRF) of the Government of India. TSM acknowledges funding from
DST/ICPS/QuST/2019/Q67.

\bibliographystyle{apsrev4-1}
\bibliography{ref.bib}

\end{document}